\documentclass[mathleft
]{an}
\usepackage{graphicx}
\usepackage{times}
\newcommand{\ov}{\overline}
\newcommand{\pdv}[2]{\frac{\partial #1}{\partial #2}}

\newcommand{\aap}{{A\&A}}		
\newcommand{\solphys}{{Sol.\ Phys.}}

\begin{document}

% The following seven commands are intended for editorial usage and should be ignored by
% the author(s).
\Pagespan{789}{}% Document's page range. 
% If second parameter is left empty, the last page is computed automatically.
\Yearpublication{2006}%
\Yearsubmission{2005}%
\Month{11}%   
\Volume{999}%  
\Issue{88}% 
% \DOI{This.is/not.aDOI}% 

\title{On the possibility of a bimodal solar dynamo}

\author{K. Petrovay$^{1,2,}$\thanks{Corresponding author:
  \email{K.Petrovay@astro.elte.hu}\newline}
%Example 
%for footnote, note the usage of the \texttt{fnmsep}
%command as separator between institute number and footnote mark} 
%\fnmsep\thanks{Corresponding author: 
%\email{E.Forgacs-Dajka@astro.elte.hu}\newline}
}
\titlerunning{On the possibility of a bimodal solar dynamo}
\authorrunning{Petrovay}
\institute{
$^1$E\"otv\"os University, Department of Astronomy, Budapest, Pf.~32, 
H-1518 Hungary\\
%Theoretical Institute for Advanced Research in Astrophysics, Dept. of Physics, 
$^2$T.I.A.R.A., Department of Physics, 
National Tsing Hua University, Hsinchu 30013, Taiwan}

\received{}
\accepted{}
\publonline{}


\abstract{A simple way to couple an interface dynamo model to a fast tachocline
model is presented, under the assumption that the dynamo saturation is due to a
quadratic process and that the effect of finite shear layer thickness on the
dynamo wave frequency is analoguous to the effect of finite water depth on
surface gravity waves. The model contains one free parameter which is fixed by
the requirement that a solution should reproduce the helioseismically
determined thickness of the tachocline. In this case it is found that, in
addition to this solution, another steady solution exists, characterized by a
four times thicker tachocline and 4--5 times weaker magnetic fields. It is
tempting to relate the existence of this second solution to the occurrence of
grand minima in solar activity.}

\maketitle

\section{Introduction}

The cyclically varying  magnetic field that gives rise to solar activity is
generally thought to arise as a consequence of an $\alpha\Omega$ dynamo
mechanism. The strong toroidal field near solar maximum is a product of the
solar differential rotation that winds up the weak general poloidal field.
Poloidal fields are then restored by an unspecified ``$\alpha$-effect'', the
physical meaning and spatial location of which has yet to be elucidated.

The shear associated with differential rotation is by far the strongest in a
thin layer below the convective zone, 
known as the tachocline. Current dynamo
models therefore invariably locate the $\Omega$-effect in the tachocline. The
site and physical nature of the $\alpha$-effect is much less clear. As
discussed in recent reviews (Petrovay 2000,
Charbonneau 2005, Solanki et al.~2006),
the two most widely duscussed alternatives are {\it interface dynamos,} where
$\alpha$ is concentrated near the bottom of the convective zone, separated from
the tachocline shear region by a thin interface only,  and {\it flux transport
dynamos,} where $\alpha$ is concentrated near the surface, and magnetic flux
transport by meridional circulation forges a link between the surface and the
tachocline. 

Solar dynamo models with a jump in the diffusivity a\-cross an interface (the
bottom of the convective zone) were first constructed by Ivanova and Ruzmaikin
(1976, 1977). Parker (1993) developed an analytical model thas has been
considered the prototype of interface dynamos ever since. Further studies of
interface dynamos include Tobias (1996), Charbonneau \& MacGregor (1997),
Markiel \& Thomas (1999), Petrovay \& Kerekes (2005). Flux transport dynamos
were constructed, among others, by Choudhuri, Sch\"ussler \& Dikpati (1995), 
Dikpati \& Charbonneau (1999), Dikpati et~al.\ (2004), Chatterjee, Nandy \&
Choudhuri (2004).

%In addition to the origin of solar activity, the observed thinness of the
%tachocline also asks for an explanation. Due to the meridional circulation, the
%latitudinal differential rotation imposed by turbulent stresses on the
%convective zone should penetrate into the deep interior on a timescale short
%compared to the age of the Sun (Spiegel \& Zahn ). To impede this, a
%strongly anisotropic (lateral) angular momentum transport mechanism is needed
%for which  magnetic stresses are the most obvious candidate. 

Like the dynamo models, tachocline models also come in two main varieties. {\it
Slow tachocline} models assume that the tachocline lies in the radiative
interior, where turbulence is insignificant, and it is pervaded by the steady
remnant magnetic field of the solar interior. If this field is limited to the
radiative interior, it may be able to confine the tachocline to its observed
low thickness. Whether or not such a limitation of the internal magnetic field
to the radiative interior is feasible, is currently a subject of debate
(Sule, R\"udiger \& Arlt 2005, Brun \& Zahn 2006).

{\it Fast tachocline} models, in contrast, assume that local instabilities
and/or convective overshoot maintain a moderate level of turbulence in the
tachocline. In this case the oscillatory magnetic field generated by the dynamo
will penetrate the tachocline and determine its dynamics. In fact, as current
dynamo models place the site of the $\Omega$-effect in the tachocline,
penetration of the dynamo field in the tachocline is a necessary condition for
those dynamo models to work. Following a suggestion by
Gilman (2000), the first detailed models of the fast tachocline
were constructed by Forg\'acs-Dajka \& Petrovay (2001, 2002) and further
developed by Forg\'acs-Dajka (2003).

In fast tachocline models the thickness of the tachocline is determined by the
intensity of the magnetic field generated by the dynamo. The thickness of the
tachocline, in turn, controls the dynamo via the shear. In this way a nonlinear
coupling is expected to arise between the dynamo and a fast tachocline. In the
following I will try to explore the nature of this nonlinear coupling, for
concreteness focusing on the case of an interface dynamo. Section 2 describes
the main features of the simple model used here, while Section 3 presents the
results. Section 4 concludes the paper.

\section{Elements of the model}

\subsection{Interface dynamo}

Parker's analytical kinematic interface dynamo (Parker 1993) consists of two
semiinfinite domains (say, the half-spaces $z<0$ and $z>0$ in an $xyz$
Cartesian frame) of diffusivity $\eta_-$ an $\eta_+$, respectively.  The $z>0$
domain is characterized by a constant shear ($\Omega=dv_y/dz$) and no
$\alpha$-effect, while the other is characterized by constant $\alpha$ and no
shear. If the
\begin{equation}  
  N=\frac{\alpha\Omega}{\eta_-\eta_+ k^3}
\end{equation} 
dynamo number is sufficiently high, a dynamo wave of form
\begin{equation} 
  B\propto \exp[ikx+(\sigma+i\tilde\omega)\eta_+k^2t]  
\end{equation}
is spontaneously excited at the interface (i.e.\ $\sigma>0$).  In the limit
$\eta_-/\eta_+\rightarrow 0$, relevant to the Sun, the dimensionless growth
rate $\sigma$ and the dimensionless frequency $\tilde\omega$ are given by the relations
\begin{eqnarray}
  N&&= \pm 8(\sigma+\textstyle{\frac 12})(\sigma+1)^{1/2}\sigma^{1/2} \\
  &&\sim \left\{
  \begin{array}{ll}
  4\sigma^{1/2} & \mbox{ if } \sigma\ll 1 \mbox{ (slightly supercritical case)} \\
  8\sigma^{2} & \mbox{ if } \sigma\gg 1 \mbox{ (strongly supercritical case)} 
  \end{array} \right. \nonumber
  \label{eq:sigmaparker}
\end{eqnarray}
and
\begin{eqnarray}
  \tilde\omega^2&&=\sigma(\sigma+1) \\ &&\sim \left\{
  \begin{array}{ll}
  \sigma & \mbox{ if } \sigma\ll 1 \mbox{ (slightly supercritical case)} \\
  \sigma^2 & \mbox{ if } \sigma\gg 1 \mbox{ (strongly supercritical case)} 
  \end{array} \right.\nonumber
  \label{eq:nuparker}
\end{eqnarray}

In what follows we will focus on the slighly supercritical limit. Indeed:
interpreting the butterfly diagram as the surface manifestation of an interface
dynamo wave we find for the frequency
\begin{equation} 
  \omega\equiv \tilde\omega\eta_+k^2=2\pi/22 \mbox{years}
  \simeq 9\cdot 10^{-9}\,\mbox{s}^{-1} .   
\end{equation}
Using $\lambda=R\pi/2$ ($R$ is the radius of the bottom of the convective zone)
we have 
\begin{equation} 
  k=4/R\simeq 8\cdot 10^{-11}\,\mbox{cm}^{-1} . \label{eq:kvalue}  
\end{equation}
With a convective zone diffusivity of $\eta_+=6\cdot 10^{12}\,$cm$^2/$s (Wang,
Nash \& Sheeley 1989, Petrovay and van Driel-Gesztelyi 1997) this yields
$\tilde\omega\simeq 0.23$, so $\sigma\simeq 0.05\ll 1$. Our assumption that the
slightly supercritical  limit holds in the solar case is therefore justified.

\subsection{Finite shear layer}

In Parker's original model, the shear layer representing the tachocline is
semi-infinite. In reality, the tachocline has some finite thickness $w$. As the
difference between the rotation rate of the radiative interior and the
equatorial rotation rate of the convective zone is fixed, the mean shear
$\Omega$ in the tachocline, and thereby also $N$ will scale inversely with
tachocline thickness: $N\propto 1/w$. It is not known how the solutions to
Parker's problem are influenced if the thickness of the shear layer is finite.
In analogy to the case of surface gravity waves over shallow water here we will
assume that the finite thickness of the shear layer reduces the squared mode
frequency $\tilde\omega^2$ by a factor $\tanh(kw)$. (At any rate, such a factor
does have the right asymptotic behavior for $w\rightarrow 0$ and
$w\rightarrow\infty$, expected on physical grounds.) In the slightly subcritical
case then $\sigma$ is also reduced by the same factor, so
\begin{equation} 
\sigma\propto N^2\tanh(kw)\propto \tanh(kw)/w^2  \label{eq:sigmaexpr1} 
\end{equation}
In the solar tachocline the angular velocity is expected to relax gradually to
the rigid rotation rate of the radiative interior, with some scale height $H$.
Defining $w$ arbitrarily as the depth where the residual rotation has been
reduced by two orders of magnitude, we have $w\sim H\ln 100$. Plugging this and
equation (\ref{eq:kvalue}) into (\ref{eq:sigmaexpr1}) we have
\begin{equation} 
\sigma\propto \tanh(4\ln 100 H/R)/H^2  \label{eq:sigmaexpr2} 
\end{equation}

Helioseismic studies  (Kosovichev 1996,
Basu \& Antia 2001) indicate that at low latitude the tachocline is
located immediately below the adiabatically stratified convective zone. The
equatorial rotation rate relaxes to the rigid rotation rate of the radiative
interior with a scale height of $H\sim5$--$10\,$Mm (cf. Forg\'acs-Dajka \& 
Petrovay 2002); for concreteness here we will take $H=7\,$Mm. This shows that
the tachocline is very thin indeed, its full thickness $w$ not exceeding a few
percents of the solar radius. Note that at higher latitudes the tachocline may
be marginally thicker, and it partly overlaps the adiabatic convective zone.

\subsection{Saturation}

Saturation of a kinematic dynamo may be brought about by a variety of effects,
such as the quenching of the $\alpha$-effect for strong magnetic fields, or
magnetic flux loss due to buoyancy. Without specifying the physical effect
responsible for it, here we will simply describe the nonlinear saturation in a
parametric form:
\begin{equation} 
\frac{\partial B}{\partial t}=\sigma B -aB^{1+\kappa}  
\end{equation}
This implies that in the saturated state ($\partial B/\partial t=0$)
the field strength is
\begin{equation} 
B_s^\kappa=\sigma/a= K\tanh(4\ln 100 H/R)/H^2 ,  \label{eq:saturated}  
\end{equation}
$K$ being an undetermined amplitude.

For the nonlinearity parameter the most plausible and widely used choice is
$\kappa=1$. (E.g.\ buoyant flux loss is expected to lead to a nonlinear term of
this form.) In what follows we will therefore focus on the case
$\kappa=1$. The influence of this choice of $\kappa$ on the solution will be
discussed at the end of Section 3 below.

Instead of using the heuristic expressions (7)--(10), a more satisfactory
approach would clearly involve a fully consistent solution of the nonlinear
dynamo problem with a finite tachocline. For certain cases such solutions were
presented e.g. in Chapter 10 of Zeldovich, Ruzmaikin and
Sokoloff (1984). However, neither of the cases considered
there ($\alpha$-effect and shear located in two infinitesimally thin layers or
in two semi-infinite layers, respectively, separated by a finite distance) does
include the effect of a finite layer thickness. An extension of those models to
the case of a finite shear layer may be a promising way to make further advance
in the problem, but it is beyond the scope of the present paper.

\subsection{Fast tachocline}

The relation between the dynamo field amplitude and the tachocline scale height
was derived by Forg\'acs-Dajka \& Petrovay (2001). Let us recall the main
points here. For the derivation we consider a plane parallel layer of
incompressible fluid of density $\rho$, where the viscosity $\nu$ and the
magnetic diffusivity $\eta$ are taken to be constant. At $z=0$ a periodic
horizontal shearing flow is imposed in the $y$ direction:
\begin{equation} 
  v_{y0}= v_0 \cos(kx) 
\end{equation}
(so that $x$ will correspond to heliographic latitude, while $y$ to the
longitude). We assume a two-dimensional flow pattern ($\partial/\partial y=0$)
and $v_x=v_z=0$. An oscillatory horizontal  field is prescribed in the $x$
direction as
\begin{equation} 
   B_x=B_p \cos(\omega t) 
\end{equation}
The evolution of the azimuthal components of the velocity and the magnetic
field is then described by the corresponding components of the equations of
motion and induction, respectively. Introducing $v=v_y$ and using Alfv\'en
speed units for the magnetic field
\begin{equation} 
  V_p=B_p(4\pi\rho)^{-1/2} \qquad b= B_y(4\pi\rho)^{-1/2} , 
\end{equation}
these can be written as
\begin{equation} 
  \pdv vt =V_p \cos(\omega t)\pdv bx+\nu\nabla^2 v  \label{eq:veq} 
\end{equation}
\begin{equation} 
  \pdv bt =V_p \cos(\omega t)\pdv vx+\eta\nabla^2 b \label{eq:beq} 
\end{equation}
Solutions may be sought in the form
\begin{equation} 
  v=\ov v(x, z) +v'(x,z) f(\omega t) 
\end{equation}
\begin{equation} 
  b=b'(x,z) f(\omega t+\phi) 
\end{equation}
where $f$ is a $2\pi$-periodic function of zero mean and of amplitude ${\cal
O}(1)$. ($\ov a$ denotes time average of $a$, while $a'\equiv a-\ov a$.)

The (temporal) average of equation (\ref{eq:veq}) reads
\begin{equation} 0=V_p \ov{\cos(\omega t)f(\omega t+\phi)} \partial_x b' -\nu\nabla^2 \ov v 
   \label{eq:vmean}  
\end{equation}
Subtracting this from equation (\ref{eq:veq}) yields
\begin{equation} 
   \pdv{v'}t=V_p [\cos(\omega t)f(\omega t+\phi)]'\pdv bx
    -\nu\nabla^2 v'  \label{eq:vfluc} 
\end{equation}

\begin{figure}
\includegraphics[width=\columnwidth]{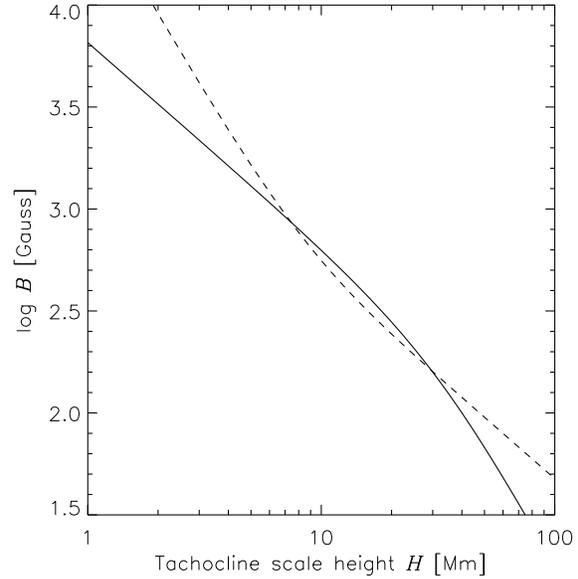}
\caption{Tachocline confining field strength $B_c$ (dashed) and dynamo
saturation field strength $B_s$ (solid) as functions of the tachocline scale
height $H$ for an interface dynamo combined with a fast tachocline. Crossing
points of the curves represent steady solutions of the nonlinear dynamo problem.
The undetermined amplitude of $B_s$ has been set so that one solution satisfies
currently available seismic constraints on $H$.}
\label{label1}
\end{figure}

For an estimate, one can suppose $\ov{\cos(\omega t)f(\omega t+\phi)}\in{\cal
O}(1)$ (i.e.\ no ``conspiracy'' between the phases, a rather natural
assumption). As $H\ll R$ we may approximate $\nabla^2\sim H^{-2}$. Estimating
the other derivatives as $\partial_t\sim\omega$ and $\partial_x\sim R^{-1}$,
(\ref{eq:vmean}) yields
\begin{equation}  
  V_p b'/R \sim \nu \ov v/H^2   \label{eq:vmeanest}  
\end{equation}
A similar order-of-magnitude estimate of the terms in equation (\ref{eq:vfluc})
yields
\begin{equation} 
  (\omega +\nu/H^2)v' \sim V_pb'/R   
\end{equation}
while from (\ref{eq:beq}) we find in a similar manner
\begin{equation} 
  \omega b'\sim (\ov{v}+v')V_p/R +\eta b'/H^2   
\end{equation}
From the last three order-of-magnitude relations one can work out with some
algebra
\begin{equation} 
   {V_p^2}= \frac{\nu R^2\omega}{H^2} 
        \frac{(1+\eta/\omega H^2)(1+\nu/\omega H^2)}{1+2\nu/\omega H^2}  
    \label{eq:estimate} 
\end{equation}
Returning from Alfvenic units to magnetic field strength we finally have
\begin{equation} 
{B_c^2}= 4\pi\rho\frac{\nu R^2\omega}{H^2} 
        \frac{(1+\eta/\omega H^2)(1+\nu/\omega H^2)}{1+2\nu/\omega H^2}  
    \label{eq:confining} 
\end{equation}

For concreteness, in the numerical evaluation of equation (\ref{eq:confining})
we will take the following fiducial values: $\nu=\eta=10^{10}\,$cm$^2/$s,
$\rho=0.08\,$g$/$cm$^3$.

\section{Results}

Setting $B=B_c=B_s$, equations (\ref{eq:saturated}) and (\ref{eq:confining})
can now be solved for $H$ and $B$. A graphical solution is presented in Figure
1. The dashed line shows relation (\ref{eq:confining}) while the solid line
represents the relation (\ref{eq:saturated}). As this relation involves a free
amplitude factor $K$, on this logarithmic plot we are free to shift the solid
curve in the vertical direction. In the figure, the solid curve was shifted so
that one crossing point falls to the value $H=7\,$Mm, the most likely value
based on helioseismic constraints. In this case the mean toroidal field
strength is seen to exceed one kilogauss (which does not exclude the existence
of $10^5\,$G flux tubes in an intermittent field structure, cf.\
Ruzmaikin 2001).

It is apparent from Figure~1 that a second stationary solution also exists,
with a roughly 4 to 5 times weaker mean magnetic field and a correspondingly
thicker tachocline. It is tempting to identify this second solution with a
``grand minimum'' state of solar activity, such as the Maunder minimum of the
17th and 18th centuries. From equations (\ref{eq:sigmaexpr2}) and
(\ref{eq:nuparker}) we expect that the dynamo period in the grand minimum mode
should be about a factor of 2--2.5 longer than in normal solar activity. There
is indeed some observational evidence indicating that during the Maunder
minimum the dominant period in solar activity may have been around 22 years
(Usoskin \& Mursula 2003), in fairly good agreement with our
expectation. (Note that the present model is plane parallel, allowing a
continuous mode spectrum. In a finite spherical geometry  the mode spectrum
will be discrete, possibly explaining why the dynamo period is exactly doubled
during grand minima.)

The results presented above are valid in the case of a quadratic nonlinearity,
i.e. $\kappa=1$. How does a change in the value of $\kappa$ influence the
validity of the findings? Analytical considerations and numerical
experimentation show that a bimodal solution persists in the range
$0.5<\kappa<2$. However, for $\kappa<0.67$ the present mode of operation would
correspond to the weak-field mode of the dynamo, which does not seem to agree
with historical evidence. On the other hand, increasing $\kappa$ above unity,
the weak-field solution becomes increasingly weaker and its tachocline depth
diverges as $\kappa\rightarrow 2$. E.g. for $\kappa=1.5$ the field strength
in the weak mode would be two orders of magnitude lower than presently, and the
tachocline thickness would increase to 300 Mm. These values are clearly
unrealistic, so the kind of bimodal solution outlined above is restricted to
saturation mechanisms whose behaviour is reasonably close to quadratic.

\section{Conclusion}

Our model contains one free parameter which is fixed by the requirement that a
solution should reproduce helioseismically derived thickness of the tachocline.
In this case we have found that, in addition to this solution, another steady
solution exists, characterized by a four times thicker tachocline and 4--5
times weaker magnetic fields. It is tempting to relate the existence this
second solution to the occurrence of grand minima in solar activity.

How can the dynamo flipflop from one of these modes to the other? Stationary
solutions of equation (\ref{eq:saturated}) are clearly linearly stable for a
fixed value of $H$. Similarly, considering the full time-dependent angular
momentum equation Forg\'acs-Dajka \& Petrovay (2002) found that the fast
tacho\-cline solution is an attractor, i.e.\ the solution of equation
(\ref{eq:confining}) for $H$ is also stable if $B_c$ is kept fixed. A more
general stability analysis of the problem in two variables is yet to be done,
but we may plausibly expect both solutions to be linearly stable if these
stationary states are to be realized in the Sun for any significant period of
time. Switching between the two solutions should then be due to finite amplitude
stochastic disturbances. On the other hand, the possibility of a more complex
nonlinear dynamical behaviour of the time-dependent system can not be discarded,
with deterministic evolution between two quasi-stationary states.

The present model clearly hinges on four main assumptions. The dynamo is
supposed to operate in an interface wave dominated regime (as opposed to a 
circulation dominated regime); the saturation mechanism is assumed to be
quadratic; the effect of finite shear layer thickness on the frequency is
assumed to be the same as for surface gravity waves over shallow water; and
finally, the tachocline is assumed to be confined by the dynamo field. The
validity of these assumptions still awaits confirmation.

\acknowledgements This research was supported by the Theoretical Institute for
Advanced Research in Astrophysics (TIARA) operated under Academia Sinica and the
National Science Council Excellence Projects program in Taiwan administered
through grant number NSC95-2752-M-007-006-PAE, as well as by the Hungarian
Science Research Fund (OTKA) under grant no.\ K67746 and by the European
Commission through the SOLAIRE Network (MTRN-CT-2006-035484).

\end{document}